%% file: bountyhunter.tex
\documentclass[acmlarge]{acmart}

\usepackage{xurl}
\usepackage{enumitem}
\usepackage{tabularx}
\usepackage{booktabs}
\usepackage{graphicx}
\usepackage{listings}
\usepackage{algorithm, algorithmicx, algpseudocode}
\usepackage{subcaption}
\usepackage{multirow}
\usepackage[frozencache,cachedir=_minted-bountyhunter]{minted}

\definecolor{LightGray}{gray}{0.95}

\begin{document}

\newcommand{\name}{Bounty Hunter}
\title{\name{}: Autonomous, Comprehensive Emulation of Multi-Faceted Adversaries}

\author{Louis Hackländer-Jansen}
\affiliation{%
  \institution{Fraunhofer FKIE}
  \city{Bonn}
  \country{Germany}}
\orcid{0009-0005-0154-9139}
\email{louis.hacklaender-jansen@fkie.fraunhofer.de}

\author{Rafael Uetz}
\affiliation{%
  \institution{Fraunhofer FKIE}
  \city{Bonn}
  \country{Germany}}
\email{rafael.uetz@fkie.fraunhofer.de}

\author{Martin Henze}
\affiliation{%
  \institution{RWTH Aachen University}
  \city{Aachen}
  \country{Germany}}
\affiliation{%
  \institution{Fraunhofer FKIE}
  \city{Bonn}
  \country{Germany}}
\email{henze@spice.rwth-aachen.de}



\input{sections/00-abstract/abstract}

\maketitle

\input{sections/10-introduction/introduction}

\input{sections/20-adversary-emulation-combined/adversary-emulation}

\input{sections/30-bounty-hunter/design-and-implementation}

\input{sections/40-evaluation/evaluation}

\input{sections/50-discussion/discussion}

\input{sections/60-ethical-concerns/ethical-concerns}

\input{sections/90-conclusion/conclusion}

\bibliography{bountyhunter}
\bibliographystyle{ACM-Reference-Format}

\input{sections/95-appendix/appendix}

\end{document}

%% file: sections/00-abstract/abstract.tex
\begin{abstract}

Adversary emulation is an essential procedure for cybersecurity assessments such as evaluating an organization's security posture or facilitating structured training and research in dedicated environments.
To allow for systematic and time-efficient assessments, several approaches from academia and industry have worked towards the automation of adversarial actions.
However, they exhibit significant limitations regarding autonomy, tactics coverage, and real-world applicability.
Consequently, adversary emulation remains a predominantly manual task requiring substantial human effort and security expertise---even amidst the rise of large language models.
In this paper, we present \name{}, an automated adversary emulation method, designed and implemented as an open-source plugin for the popular adversary emulation platform Caldera, that enables autonomous emulation of adversaries with multi-faceted behavior while providing a wide coverage of tactics.
To this end, it realizes diverse adversarial behavior, such as different levels of detectability and varying attack paths across repeated emulations.
By autonomously compromising a simulated enterprise network, \name{} showcases its ability to achieve given objectives without prior knowledge of its target, including pre-compromise, initial compromise, and post-compromise attack tactics.
Overall, \name{} facilitates autonomous, comprehensive, and multi-faceted adversary emulation to help researchers and practitioners in performing realistic and time-efficient security assessments as well as research and training in intrusion detection, incident response, and forensic analysis.

\end{abstract}

%% file: sections/10-introduction/introduction.tex
\section{Introduction}

\emph{Adversary emulation} is an integral component of security assessments, training, and research in both industry and academia, complementing related methods such as penetration testing and vulnerability scanning~\cite{landauerRedTeamRedemption2024, applebaumANALYSISAUTOMATEDADVERSARY2017, applebaumIntelligentAutomatedRed2016, millerAutomatedAdversaryEmulation2018}.
It assesses the overall security of a target system or network by emulating a full attack lifecycle of real-world attackers, utilizing their tactics, techniques, and procedures (TTPs) to evaluate the target's security readiness, mechanisms, and countermeasures~\cite{zilbermanSoKSurveyOpenSource2020,applebaumIntelligentAutomatedRed2016,millerAutomatedAdversaryEmulation2018, applebaumANALYSISAUTOMATEDADVERSARY2017}.
Furthermore, it allows structured testing and research of security mechanisms~\cite{debattyBuildingCyberRange2019, parkDesignImplementationMultiCyber2022, leitnerAITCyberRange2021} as well as training through security exercises~\cite{davisSurveyCyberRanges2013, vykopalLessonsLearnedComplex2017, kovacevicRedTeamsPentesters2020a, landauerRedTeamRedemption2024}.
However, \emph{manual} adversary emulation is time-con\-sum\-ing, offers limited reproducibility, and requires considerable expertise~\cite{applebaumIntelligentAutomatedRed2016, applebaumANALYSISAUTOMATEDADVERSARY2017, millerAutomatedAdversaryEmulation2018, zilbermanSoKSurveyOpenSource2020}. %
To address these issues, various works have proposed \emph{automated attack planning methods} wherein attack paths, i.e., sequences of adversarial actions, are automatically generated to achieve a given goal~\cite{hoffmannSimulatedPenetrationTesting2015, wotawaPlanItAutomated2014, yooCyberAttackDefense2020, sarrautePOMDPsMakeBetter2013}.
Yet, these approaches fall short of showing their real-world applicability through actual implementations of realistic attack sequences and are often limited to highly specific scenarios.
More recent works have explored how large language models (LLMs) can be employed for automated security assessments~\cite{dengPentestGPtEvaluatingHarnessing, xuAutoAttackerLargeLanguage2024, wangSandsMansionsAutomated2025, happeLLMsHackersAutonomous2024}.
While they present promising results, they generally struggle to perform high-level, multi-step action planning, thus failing to emulate attack scenarios in a fully autonomous manner.
Additionally, LLM-based approaches inherit several limitations from LLMs, such as hallucinations~\cite{xuAutoAttackerLargeLanguage2024, dengPentestGPtEvaluatingHarnessing, bangMultitaskMultilingualMultimodal2023, jiSurveyHallucinationNatural2023} and LLM-related security concerns~\cite{vassilevAdversarialMachineLearning2024, owaspgenaiprojectGenAIIncidentResponse2025}.
On the practical side, \emph{adversary emulation tools} can automate certain tasks, such as executing malicious scripts and exploiting vulnerabilities~\cite{landauerRedTeamRedemption2024}.
However, they do not provide autonomy, a wide coverage of TTPs, and adaptable adversarial attributes, which are requirements identified as important for adversary emulation~\cite{millerAutomatedAdversaryEmulation2018, pruzinecKUBOFrameworkAutomated2022, zilbermanSoKSurveyOpenSource2020, landauerRedTeamRedemption2024} and adversarial activity in testing and training environments~\cite{uetzReproducibleAdaptableLog2021, vykopalLessonsLearnedComplex2017, davisSurveyCyberRanges2013}.

To overcome the above limitations in the domain of adversary emulation, we present \name{}, a novel online planning method that allows for an autonomous emulation of adversaries with multi-faceted behavior that perform techniques from pre-compromise, initial compromise, and post-compromise tactics.
It links separate actions using their pre- and post-conditions and employs probabilistic, reward-driven decision making to generate varying attack paths towards user-defined goals.
By employing certain characteristics of actions (e.g., their detectability) during its reward calculation, \name{} can emulate adversaries with varying desired attributes (e.g., their level of stealthiness).

To demonstrate and validate our method, we implemented \name{} as a plugin for the prominent adversary emulation platform Caldera.
As a proof of concept, we show that it can autonomously achieve its goals without prior knowledge of the target system, given a set of attack actions implementing suitable techniques (e.g., the attack actions provided by Caldera).
To this end, we simulate several multi-step attacks as performed by real-world threat groups, e.g., gaining initial access to a Windows Active Directory domain, escalating privileges, and compromising the entire domain---thus covering a wide range of tactics.
We also show that it facilitates variability in adversarial behavior, such as different levels of detectability and varying attack sequences when repeating a scenario.

\noindent\textbf{Contributions.}
To summarize, our contributions are:

\begin{itemize}[leftmargin=*]
    \item We identify essential requirements for adversary emulation based on related work and show that existing approaches exhibit deficiencies in autonomy, TTP coverage, and variability in adversarial behavior (§\ref{sec:adversary-emulation}).
    \item We present \name{}, the first planning method that allows for an autonomous emulation of adversaries with a wide range of tactics while exhibiting adaptable behavioral attributes and a variety in attack paths (§\ref{sec:design-and-implementation}).
    \item To spur further research and enable practical application, we implemented \name{} as an open-source plugin\footnote{\name{} is publicly available as an open-source plugin for Caldera at https://github.com/fkie-cad/bounty-hunter.
We deliberately exclude sensitive attack actions for ethical reasons (cf. Section~\ref{sec:concerns}).} for the adversary emulation platform Caldera.
    \item Our experiments show how \name{} enhances security assessments, allowing organizations to improve their security posture and providing researchers with a valuable foundation for sound research (§\ref{sec:evaluation}).
\end{itemize}


%% file: sections/20-adversary-emulation-combined/adversary-emulation.tex
\section{Adversary Emulation}
\label{sec:adversary-emulation}

As organizations increasingly rely on digital infrastructure, the risks associated with cyber threats have escalated in recent years~\cite{AllianzRiskBarometer}.
Thus, it is crucial for organizations to identify and remediate potential weak spots within their enterprise network before an attacker can exploit them and to put appropriate countermeasures as well as reaction strategies in place~\cite{mughalBuildingSecuringModern2022}.
Security assessments such as penetration testing and vulnerability scanning have established themselves as critical components to help in this area~\cite{shindeCyberSecurityAnalysis2016}.
However, they usually have a narrow scope with respect to the assessed systems and focus solely on identifying and exploiting vulnerabilities~\cite{landauerRedTeamRedemption2024, applebaumANALYSISAUTOMATEDADVERSARY2017, kovacevicRedTeamsPentesters2020a}.
Adversary emulation complements these approaches by emulating real-world adversarial tactics, techniques, and procedures (TTPs)~\cite{MitreAttackMatrix}.
It covers the complete lifecycle of an attack to fully evaluate the security posture and countermeasures of an organization or system to gain insight of what the assessed system looks like from an attacker's point of view~\cite{applebaumIntelligentAutomatedRed2016, zilbermanSoKSurveyOpenSource2020, millerAutomatedAdversaryEmulation2018}.
In addition, it can also be used in training environments (e.g., cyber ranges~\cite{davisSurveyCyberRanges2013, debattyBuildingCyberRange2019, parkDesignImplementationMultiCyber2022}), facilitating high levels of interaction between attackers and defenders.
This allows for realistic training exercises in which trainees have to defend a system against realistic threats, thus helping to test and improve detection and response capabilities of security teams~\cite{landauerRedTeamRedemption2024, kovacevicRedTeamsPentesters2020a, vykopalLessonsLearnedComplex2017, davisSurveyCyberRanges2013, zilbermanSoKSurveyOpenSource2020}.
Furthermore, adversary emulation allows to generate and reproduce simulated attacks for structured analysis and comparison of security mechanisms, enabling security researchers and vendors to evaluate security mechanisms as well as cybersecurity products against known threats~\cite{zilbermanSoKSurveyOpenSource2020, millerAutomatedAdversaryEmulation2018, ATTACKEvals}.

\input{sections/20-adversary-emulation-combined/aae}

\input{sections/20-adversary-emulation-combined/reqs}

\input{sections/20-adversary-emulation-combined/current-state}

%% file: sections/20-adversary-emulation-combined/aae.tex
\subsection{Automated Adversary Emulation}
\label{subsec:aae}

Since manual adversary emulation assessments are time-consuming, require extensive expertise, and lack reproducibility~\cite{zilbermanSoKSurveyOpenSource2020, applebaumANALYSISAUTOMATEDADVERSARY2017, applebaumIntelligentAutomatedRed2016, millerAutomatedAdversaryEmulation2018}, several researchers and practitioners have worked towards automating adversary emulation.
To motivate the need for our work, we first exemplify the general process of automated adversary emulation alongside the most prominent adversary emulation tool \emph{Caldera}~\cite{Caldera, applebaumIntelligentAutomatedRed2016, landauerRedTeamRedemption2024} and its state-of-the-art decision logic, the \emph{Look Ahead Planner}~\cite{CalderaLookAheadPlanner}.
Caldera encompasses over 1,700 attack actions, choosing and executing them in a sequential, autonomous manner.
Before an assessment, users can define reward values for certain adversarial actions (e.g., \textit{Exfiltrate Sensitive Files}).
The defined reward values are used to calculate the \textit{anticipated future rewards} of all available actions. 
Leveraging these future rewards, the Look Ahead Planner then executes action sequences to maximize its reward, autonomously gathering information required for the execution of high-reward actions (e.g., \textit{Find}, \textit{Stage}, and \textit{Exfiltrate Sensitive Files}).

However, as we discuss in Section~\ref{subsec:related-work}, many approaches (including Caldera) only emulate post-compromise techniques, neglecting the initial compromise of a target system, or focus on highly specific scenarios, disregarding crucial attack tactics and techniques for comprehensive security assessments and trainings.
Furthermore, repeated automated assessments generally result in identical attack sequences, which is particularly unfavorable for recurring training exercises.
Finally, most approaches do not address configuring the emulated adversary's behavior, e.g., varying difficulties for training exercises or emulating designated threat actors for benchmarking security mechanisms.

%% file: sections/20-adversary-emulation-combined/reqs.tex
\subsection{Requirements for Adversary Emulation}
\label{subsec:requirements}

Various works have analyzed the challenges of adversary emulation and proposed desirable properties in general as well as for its application in testbeds and cyber ranges for research and training purposes.
We give an overview of these works to identify desirable properties and establish \emph{four essential requirements} for adversary emulation.

Landauer et al.~\cite{landauerRedTeamRedemption2024} interviewed 20 cybersecurity professionals using 80 questions on adversary emulation tools, finding that automation, configurability, ease of use, and a wide coverage of TTPs are among the most-desired properties.
Zilberman et al. ~\cite{zilbermanSoKSurveyOpenSource2020} present similar criteria, which they use to conduct a survey of prominent adversary emulation tools.
Moving from surveys and reviews to emulation approaches, Miller et al.~\cite{millerAutomatedAdversaryEmulation2018} define autonomous execution, resilience regarding uncertainty, realism, ease of use, and a modular design as key properties for automated adversary emulation.
Pružinec et al.~\cite{pruzinecKUBOFrameworkAutomated2022} present a framework for testing malware detection (KUBO), and postulate that emulated attacks should be semantically coherent in their sequences, i.e., separate attack actions should run in each other's context and not in the emulator's context, particularly during initial access, privilege escalation, and lateral movement.

From a more general perspective, different works derive desired properties for cyber ranges and testbeds.
Davis and Magrath~\cite{davisSurveyCyberRanges2013} review, categorize, and discuss cyber ranges and testbeds, concluding that their main advantage is enabling training exercises with varying complexity and difficulty as well as repeatable scenarios.
Drawing from practical experience, Vykopal et al.~\cite{vykopalLessonsLearnedComplex2017} present their greatest challenges during cyber range trainings as varying the exercise difficulties to match the proficiency of trainees, automating attacks, and handling uncertainties during attack execution.
Uetz et al.~\cite{uetzReproducibleAdaptableLog2021} define properties for sound and reproducible cybersecurity experiments and testbeds in general, finding that they should be realistic, adaptable, and reproducible.
Regarding training in general, Christina and Bjork~\cite{christinaOptimizingLongtermRetention1991} state that training scenarios should exhibit variability and variety to enhance transfer of knowledge.

Based on the above findings, we derive \emph{four essential requirements} for adversary emulation approaches:

\noindent\textbf{(R1) Autonomous Goal Pursuit.}
The emulated adversary should au\-to\-no\-mous\-ly decide which actions to perform to reach its user-defined goal~\cite{millerAutomatedAdversaryEmulation2018, vykopalLessonsLearnedComplex2017, xuAutoAttackerLargeLanguage2024}.
This includes information gathering required for further execution and handling uncertainties occurring during emulation (e.g., failing actions), without human interaction or predefined sequences.
Users can thus run assessments by defining one or more goals, e.g., \textit{Exfiltrate Sensitive Files} without defining playbooks or target details such as infrastructure or filenames.

\noindent\textbf{(R2) Wide Coverage of TTPs.}
The emulated adversary should cover a wide range of tactics, techniques, and procedures for comprehensive security assessments~\cite{zilbermanSoKSurveyOpenSource2020, landauerRedTeamRedemption2024, millerAutomatedAdversaryEmulation2018, dengPentestGPtEvaluatingHarnessing, wangSandsMansionsAutomated2025}.
The MITRE ATT\&CK framework~\cite{MitreAttackMatrix} classifies adversarial techniques into 15 tactics: two pre-compromise (\textit{Reconnaissance}, \textit{Resource Development}), one initial compromise (\textit{Initial Access}), and 12 post-compromise.
Furthermore, the covered techniques should be emulated \textit{coherently}, e.g., after privilege escalation, subsequent commands should run in the elevated context instead of the emulator's original context~\cite{pruzinecKUBOFrameworkAutomated2022}.

\noindent\textbf{(R3) Adaptable Adversarial Attributes.}
The emulated adversary should be adaptable in its attributes to reflect different behaviors.
For example, for usage in security exercises and trainings, the proficiency of the adversary should be adaptable to allow for varying levels of difficulty~\cite{vykopalLessonsLearnedComplex2017, davisSurveyCyberRanges2013}.
Additionally, when assessing and benchmarking security mechanisms, restricting an emulation to techniques seen as part of a real campaign from a designated threat actor allows for structured experiments and tests~\cite{ATTACKEvals, wangSandsMansionsAutomated2025}.

\noindent\textbf{(R4) Attack Path Variety Across Scenario Repetition.}
When a scenario is re-run, the emulated adversary should produce varied attack paths, so that defenders who already practiced the scenario cannot trivially detect the attack~\cite{christinaOptimizingLongtermRetention1991, wangSandsMansionsAutomated2025}.
At the same time, it must be possible to repeat a scenario with identical attack sequences (e.g., via a randomization seed) to enable sound comparisons of security mechanisms and configurations~\cite{uetzReproducibleAdaptableLog2021, davisSurveyCyberRanges2013}.

%% file: sections/20-adversary-emulation-combined/current-state.tex
\subsection{Current State of Adversary Emulation}
\label{subsec:related-work}

To allow for accessible, cost-efficient, and repeatable assessments, various related work on automating adversarial actions has been published in academia and industry~\cite{hoffmannSimulatedPenetrationTesting2015, landauerRedTeamRedemption2024, zilbermanSoKSurveyOpenSource2020}.
We give an overview of academic and practical approaches and assess them against our four essential requirements (cf. Table~\ref{tab:current-state}).

\input{sections/20-adversary-emulation-combined/table-current-state}

Academic research has primarily focused on automated planning methods, wherein attacks are described as a sequence of actions to achieve a given goal~\cite{applebaumIntelligentAutomatedRed2016}.
A foundational planning system is STRIPS~\cite{fikesStripsNewApproach1971}, which defines conditions under which each action can be executed (pre-condition) and the effects of each action on the current state (post-condition).
Wotawa and Bozic~\cite{wotawaPlanItAutomated2014} transfer this concept to attack planning for automated security testing of web applications.
Yoo et al.~\cite{yooCyberAttackDefense2020} use a STRIPS-based approach to generate multiple attack scenarios, implementing both pre- and post-compromise tactics, but only eight different techniques.
Sarraute et al.~\cite{sarrautePOMDPsMakeBetter2013} propose a model employing partially observable Markov decision processes~\cite{kaelblingPlanningActingPartially1998} to account for uncertainties in network configurations for specific penetration testing scenarios.
Notably, all these methods perform offline planning, i.e., they generate action sequences before execution and therefore cannot adapt to new information or uncertainties discovered during emulation~\cite{wangRedTeamAutomated2024}.
To overcome this limitation, Applebaum et al.~\cite{applebaumIntelligentAutomatedRed2016} present Caldera, an adversary emulation tool that plans with actions' conditions while reacting to unexpected results at emulation-time.
We further describe this practical tool below.
Based on this work, Wang et al.~\cite{wangRedTeamAutomated2024} propose a model that adapts its strategy based on execution results and ensures a given order of actions, even when they are not logically linked via conditions.
Both approaches are limited to post-compromise actions and therefore lack an emulation of (pre-)compromise, i.e., pre-compromise and initial compromise, actions of the attack lifecycle.
In line with prior work~\cite{applebaumIntelligentAutomatedRed2016, wangRedTeamAutomated2024,millerAutomatedAdversaryEmulation2018}, we conclude that adversary emulation is inherently both a planning and acting problem, rendering offline planners inappropriate~\cite{millerAutomatedAdversaryEmulation2018,  applebaumANALYSISAUTOMATEDADVERSARY2017, applebaumIntelligentAutomatedRed2016, wangRedTeamAutomated2024, alfordCALDERARedBlueCyber2022}.
In general, the reviewed academic approaches tend to lack autonomous goal pursuit, tactic coverage, and real-world applicability~\cite{wangRedTeamAutomated2024, applebaumIntelligentAutomatedRed2016, millerAutomatedAdversaryEmulation2018, applebaumANALYSISAUTOMATEDADVERSARY2017}.

More recent studies investigate how large language models (LLMs) can support offensive security and penetration testing~\cite{guptaChatGPTThreatGPTImpact2023, mohamedfirdhousWormGPTLargeLanguage2023}.
Deng et al.~\cite{dengPentestGPtEvaluatingHarnessing} propose PentestGPT, an LLM-empowered penetration testing framework that iteratively proposes actions that users execute and return feedback.
Xu et al.~\cite{xuAutoAttackerLargeLanguage2024} propose AutoAttacker, an LLM-guided system for automated simulation of attacks.
However, it lacks an understanding of actions' conditions, limiting its autonomy in complex multi-step scenarios.
Wang et al.~\cite{wangSandsMansionsAutomated2025} propose AURORA, which generates attack chains from threat intelligence reports and tool documentation using pre- and post-conditions and executes them semi-automatically.
While the authors mention that AURORA supports variety in its generated attack chains, reproducibility and diversity control are not clearly described.
Happe et al.~\cite{happeLLMsHackersAutonomous2024} propose their LLM-guided prototype wintermute, which autonomously emulates privilege escalation techniques but does not cover other tactics.
To conclude, LLM-based approaches struggle to robustly model and plan with actions' pre- and post-conditions, limiting fully autonomous multi-step attack planning and execution.
They also inherit model-related issues, such as hallucinations~\cite{xuAutoAttackerLargeLanguage2024, dengPentestGPtEvaluatingHarnessing, bangMultitaskMultilingualMultimodal2023, jiSurveyHallucinationNatural2023} and LLM-related security concerns~\cite{vassilevAdversarialMachineLearning2024, owaspgenaiprojectGenAIIncidentResponse2025}.
The generated sequences heavily depend on the knowledge of the employed model, leading to potentially outdated vulnerability information and biased attack preferences (e.g., prioritizing brute-force methods)~\cite{dengPentestGPtEvaluatingHarnessing, happeLLMsHackersAutonomous2024, xuAutoAttackerLargeLanguage2024}.

On the more practical side, security personnel often employ adversary emulation tools to (partially) automate adversarial actions~\cite{landauerRedTeamRedemption2024, zilbermanSoKSurveyOpenSource2020}.
These include the execution of malicious commands or scripts to automate common actions, such as enumerating system information, exploiting known vulnerabilities, or dumping credentials.
The implementations of such practical tools range from straightforward scripts that execute single malicious commands to complex systems, comprising command-and-control (C2) servers and agents that can execute predefined attack sequences.
We surveyed the current landscape of adversary emulation tools and agree with Landauer et al.~\cite{landauerRedTeamRedemption2024}, who deemed the five following tools \textit{most suitable} in their 2024 survey (Zilberman et al.~\cite{zilbermanSoKSurveyOpenSource2020} came to similar conclusions). %
APTSimulator~\cite{APTSimulator} offers straightforward Windows batch scripts that simulate realistic threats on Windows targets, mapping actions to MITRE ATT\&CK~\cite{MitreAttackMatrix} but with limited tactic coverage.
Atomic Red Team~\cite{AtomicRedTeam} is a library of portable adversarial actions covering 14 of 15 ATT\&CK tactics, only missing \textit{Resource Development}~\cite{atomicredteam-coverage}.
Neither tool includes a planning component.
Caldera~\cite{Caldera} is an automated adversary emulation platform with a C2 infrastructure and explicit support for custom planning methods, that can employ its extensible library of over 1{,}700 post-compromise attack actions~\cite{alfordCALDERARedBlueCyber2022}.
Infection Monkey~\cite{InfectionMonkey} focuses on network security assessments and simulates worm-like propagation emphasizing initial access and lateral movement.
Metasploit~\cite{Metasploit} is a widely used penetration testing framework that discovers and exploits vulnerabilities and deploys payload-carrying modules to obtain interactive post-compromise shells.
Across these tools, no single solution satisfies a broad set of our essential requirements (cf.\ Table~\ref{tab:current-state}): None combines autonomy with wide TTP coverage, and only AURORA addresses adaptable behavior and attack path variety.
Nonetheless, we agree with Landauer et al.~\cite{landauerRedTeamRedemption2024} that Caldera is currently the most suitable platform for practical use, as its design explicitly supports integrating novel planning methods and thereby helps bridge the gap between theoretical planning research and practical adversary emulation~\cite{alfordCALDERARedBlueCyber2022}.

%% file: sections/20-adversary-emulation-combined/table-current-state.tex
\begin{table}
\small
\centering
\setlength\tabcolsep{2mm}
\caption{Our review of academic work on attack planning and practical adversary emulation tools identifies substantial limitations regarding the four essential requirements we established based on related work on adversary emulation.}
\begin{tabularx}{0.95\linewidth}{
    p{3pt}
    p{80pt}
    >{\centering\arraybackslash}p{67pt}
    >{\centering\arraybackslash}p{67pt}
    >{\centering\arraybackslash}p{76pt}
    >{\centering\arraybackslash}p{67pt}
}
\toprule
    & & Autonomy & TTP Coverage & Adaptable Attributes & Path Variety \\
\midrule
    \multirow{8}{*}{\rotatebox{90}{Academic}}
    & Wotawa \& Bozic~\cite{wotawaPlanItAutomated2014} &
      \includegraphics{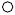} &
      \includegraphics{images/circles/circle-empty.pdf} &
      \includegraphics{images/circles/circle-empty.pdf} &
      \includegraphics{images/circles/circle-empty.pdf} \\
    & Yoo et al.~\cite{yooCyberAttackDefense2020} &
      \includegraphics{images/circles/circle-empty.pdf} &
      \includegraphics{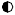} &
      \includegraphics{images/circles/circle-empty.pdf} &
      \includegraphics{images/circles/circle-empty.pdf} \\
    & Sarraute et al.~\cite{sarrautePOMDPsMakeBetter2013} &
      \includegraphics{images/circles/circle-empty.pdf} &
      \includegraphics{images/circles/circle-empty.pdf} &
      \includegraphics{images/circles/circle-empty.pdf} &
      \includegraphics{images/circles/circle-empty.pdf} \\
    & Wang et al.~\cite{wangRedTeamAutomated2024} &
      \includegraphics{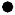} &
      \includegraphics{images/circles/circle-empty.pdf} &
      \includegraphics{images/circles/circle-empty.pdf} &
      \includegraphics{images/circles/circle-empty.pdf} \\
    & Deng et al.~\cite{dengPentestGPtEvaluatingHarnessing} &
      \includegraphics{images/circles/circle-empty.pdf} &
      \includegraphics{images/circles/circle-full.pdf} &
      \includegraphics{images/circles/circle-empty.pdf} &
      \includegraphics{images/circles/circle-empty.pdf} \\
    & Xu et al.~\cite{xuAutoAttackerLargeLanguage2024} &
      \includegraphics{images/circles/circle-empty.pdf} &
      \includegraphics{images/circles/circle-full.pdf} &
      \includegraphics{images/circles/circle-empty.pdf} &
      \includegraphics{images/circles/circle-empty.pdf} \\
    & Wang et al.~\cite{wangSandsMansionsAutomated2025} &
      \includegraphics{images/circles/circle-empty.pdf} &
      \includegraphics{images/circles/circle-full.pdf} &
      \includegraphics{images/circles/circle-half.pdf} &
      \includegraphics{images/circles/circle-half.pdf} \\
    & Happe et al.~\cite{happeLLMsHackersAutonomous2024} &
      \includegraphics{images/circles/circle-full.pdf} &
      \includegraphics{images/circles/circle-empty.pdf} &
      \includegraphics{images/circles/circle-empty.pdf} &
      \includegraphics{images/circles/circle-empty.pdf} \\

\midrule

\multirow{5}{*}{\rotatebox{90}{Practical}}
    & APTSimulator~\cite{APTSimulator} &
      \includegraphics{images/circles/circle-empty.pdf} &
      \includegraphics{images/circles/circle-empty.pdf} &
      \includegraphics{images/circles/circle-empty.pdf} &
      \includegraphics{images/circles/circle-empty.pdf} \\
    & Atomic Red Team~\cite{AtomicRedTeam} &
      \includegraphics{images/circles/circle-empty.pdf} &
      \includegraphics{images/circles/circle-full.pdf} &
      \includegraphics{images/circles/circle-empty.pdf} &
      \includegraphics{images/circles/circle-empty.pdf} \\
    & Caldera~\cite{Caldera}&
      \includegraphics{images/circles/circle-full.pdf} &
      \includegraphics{images/circles/circle-empty.pdf} &
      \includegraphics{images/circles/circle-empty.pdf} &
      \includegraphics{images/circles/circle-empty.pdf} \\
    & Infection Monkey~\cite{InfectionMonkey} &
      \includegraphics{images/circles/circle-full.pdf} &
      \includegraphics{images/circles/circle-empty.pdf} &
      \includegraphics{images/circles/circle-empty.pdf} &
      \includegraphics{images/circles/circle-empty.pdf} \\
    & Metasploit~\cite{Metasploit} &
      \includegraphics{images/circles/circle-empty.pdf} &
      \includegraphics{images/circles/circle-full.pdf} &
      \includegraphics{images/circles/circle-empty.pdf} &
      \includegraphics{images/circles/circle-empty.pdf} \\

\midrule

    & \name{} &
      \includegraphics{images/circles/circle-full.pdf} &
      \includegraphics{images/circles/circle-full.pdf} &
      \includegraphics{images/circles/circle-full.pdf} &
      \includegraphics{images/circles/circle-full.pdf} \\

\bottomrule

\end{tabularx}
\label{tab:current-state}
\end{table}

%% file: sections/30-bounty-hunter/design-and-implementation.tex
\section{\name{}}
\label{sec:design-and-implementation}

To advance approaches in the field of adversary emulation and to enhance security assessments, training exercises, and intrusion detection research, we present \name{}, the first planning method for the autonomous emulation of adversaries with multi-faceted behaviors that cover pre-, initial, and post-compromise tactics.
In its core, \name{} uses the well-established method of constructing links between actions using their pre- and post-conditions and leveraging them to autonomously pursue its user-defined goals driven by their reward values.
It substantially expands upon related approaches by employing reward calculation based on specific properties of actions, which enables the emulation of adversaries with desired attributes (e.g., stealthy vs. easy-to-detect adversaries).
Additionally, it employs probabilistic action selection to allow varying attack paths when repeating the same assessment multiple times.
To summarize, \name{}'s key features are:

\begin{enumerate}[label=(\arabic*), topsep=0pt, itemsep=0pt]
    \item autonomous, reward-driven planning,
    \item pre-compromise, initial compromise, and privilege escalation coverage,
    \item reward calculation based on action properties, and
    \item probabilistic action selection.
\end{enumerate}

\begin{figure}
    \centering
    \includegraphics[width=0.8\linewidth]{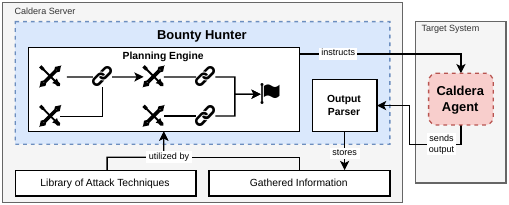}
    \caption{
      We designed \name{} as a plugin for Caldera to adopt its library of over 1,700 attack techniques as well as management capabilities of agents and their gathered information.
      \name{}'s planning engine links actions using their respective pre- and post-conditions to construct attack sequences to reach its user-defined goal.
    }
    \label{fig:bh-caldera}
\end{figure}

Features (1) and (2) are also implemented by other approaches and therefore not entirely novel from a scientific perspective.
However, other approaches do not implement both features at the same time (cf. Table~\ref{tab:current-state}), which is crucial for security assessments (cf. Section~\ref{subsec:requirements}).
While we acknowledge that features (3) and (4) are partially addressed by AURORA~\cite{wangSandsMansionsAutomated2025}, \name{} is the first approach to implement them as required for practical application (cf. Section~\ref{sec:evaluation}).

We designed \name{} as a plugin for Caldera (cf. Figure~\ref{fig:bh-caldera}), whose modular design allows for custom planning methods to adopt its library of over 1,700 attack actions and its capabilities (e.g., managing agents and their gathered information)~\cite{alfordCALDERARedBlueCyber2022}.
\name{} uses Caldera's command-and-control (C2) infrastructure, comprising a C2 server that manages the planning process and C2 agents that execute attack actions on the target systems.
Employing Caldera's execution methodology, a designated planning engine instructs the agents when and in what order they should perform individual attack actions from the set of provided techniques.
In the following, we describe the design of \name{} regarding its key features in more detail.

\subsection{Autonomous, Reward-Driven Planning}
\label{sec:planningmethod}

To keep the entry barrier low and allow for time-efficient assessment preparation, users should not have to provide pre-written playbooks or detailed information about the target system, such as target infrastructure or filenames (cf. Section~\ref{subsec:requirements}).
\name{} tries to autonomously reach its user-defined goal by utilizing the well-established method of leveraging pre- and post-conditions of actions and reward-driven decision making~\cite{fikesStripsNewApproach1971, CalderaLookAheadPlanner, wotawaPlanItAutomated2014, yooCyberAttackDefense2020, wangRedTeamAutomated2024, wangSandsMansionsAutomated2025}.
Advancing related approaches, it provides options for automatically (un)locking actions and updating rewards during a running assessment.
This allows for an emulation of adversaries with intermediate goals (e.g., establishing persistence before exfiltrating files) as well as ensuring a specific order of execution when actions are not logically linked via pre- and post-conditions~\cite{wangRedTeamAutomated2024}.
In the following, we give a formal definition of \name{}'s planning model and describe its design towards its autonomous, reward-driven planning.

\textbf{Formal Definition of the Planning Model.}
We define our planning problem as \(\mathcal{P} = (\mathcal{K}, \mathcal{A}, \mathcal{G})\).
\(\mathcal{K}\) is the current knowledge, which is the set of information items about the target system.
The action set \(\mathcal{A}\) defines attack actions that can be executed, where an action \(a \in \mathcal{A}\) is defined as \(a = (r_a, \mathcal{C}^{pre}_a, \mathcal{C}^{post}_a)\).
\(r_a\) is the reward of the action, \(\mathcal{C}^{pre}_a\) is the set of pre-conditions of the action, i.e., information items required in \(\mathcal{K}\) for the execution of \(a\), and \(\mathcal{C}^{post}_a\) is the set of post-conditions of the action, i.e., information items gathered and stored in \(\mathcal{K}\) by the successful execution of \(a\).
Executing an action \(a\) results in a transition to a new state, where the model's knowledge is expanded by the action's post-conditions (\(\mathcal{K}' = \mathcal{K} \cup \mathcal{C}^{post}_a\)).
An action \(a\) is \emph{executable} if and only if all its pre-conditions are fulfilled (\(\forall c \in \mathcal{C}^{pre}_a: c \in \mathcal{K}\)).
An action \(a_2\) is a \emph{following} action of action \(a_1\) if and only if \(a_2\) has a pre-condition that is a post-condition of \(a_1\) (\(\exists k: k \in \mathcal{C}^{pre}_{a_2}, k \in \mathcal{C}^{post}_{a_1}\)).
The goal actions \(\mathcal{G} \subseteq \mathcal{A}\) are a set of actions that the user configured as goal(s) of the emulated adversary.
A solution to \(\mathcal{P}\) is a sequence of actions \(s = (a_1, a_2, ..., a_n)\) where \(a_n \in \mathcal{G}\), i.e., a sequence that ends in the execution of a goal action.

\begin{figure*}
    \centering
    \includegraphics[width=0.8\linewidth]{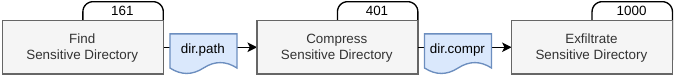}
    \caption{
      \name{} uses pre- and post-conditions to link actions and recursively calculate their future reward values.
    }
    \label{fig:linkedsteps}
\end{figure*}

\textbf{Autonomous Action Linking.}
To eliminate the need for human interaction or predefined action sequences, \name{} autonomously links attack actions to generate attack sequences by utilizing their respective pre- and post-conditions.
Most techniques in Caldera's library adopted by \name{} already come with defined pre- and post-conditions.
\name{} autonomously links two actions \((a_1, a_2)\) when action \(a_2\) is a following action of action \(a_1\) to construct possible action sequences.
Figure~\ref{fig:linkedsteps} shows such a sequence \(s=({a_1}, {a_2}, {a_3})\) with its linked actions and their respective pre- and post-conditions.
For example, the action \textit{Compress Sensitive Directory} (\(a_2\)) requires information about directories containing sensitive data (\({C}^{pre}_{a_2}=\{\mathrm{dir.path}\}\)).
The action \textit{Find Sensitive Directory} (\(a_1\)) gathers this information (\({C}^{post}_{a_1}=\{\mathrm{dir.path}\}\)).
Since \(\mathrm{dir.path} \in {C}^{pre}_{a_2}\) and \(\mathrm{dir.path} \in {C}^{post}_{a_1}\), \name{} links the two actions.

\textbf{Reward-Driven Decision Making.}
To limit the required user input to the definition of goals, \name{} leverages its links between actions in combination with reward values of actions to calculate the \textit{anticipated future reward values} for all its actions.
These future reward values are used as basis for its decision making process, wherein it sequentially executes the action with no unfulfilled pre-conditions and the highest future reward.
The base reward values of actions are assigned in two ways: (1) when a user defines actions as goals, \name{} automatically assigns these actions a high goal reward and (2) users can optionally assign custom rewards to individual actions.
\name{} recursively calculates the future reward of an attack action \(a\) using a finite-horizon discount model~\cite{kaelblingPlanningActingPartially1998} with the underlying Formula~\ref{eq:future_reward} shown below.
The future reward of each action is calculated by adding its own reward to the maximum of discounted future reward values of its following actions.
The exponential discount factor \(g^{d}\) includes the number of steps from the current action \(d\) and reduces the impact of action rewards that lie further down an attack sequence, with a configurable maximum depth that limits the number of considered actions in a sequence.
In case an action fails, \name{} automatically chooses and executes an alternative action to reach its goal, driven by the future rewards of actions that increase towards the goal.
\begin{equation}
\label{eq:future_reward}
    f(a,d) = r_a \times\ g^{d} + \max(f(a.\mathrm{following},d+1))
\end{equation}
Figure~\ref{fig:linkedsteps} shows how \name{} finds a solution for a planning problem \(\mathcal{P}\) using its calculated future rewards.
We define \textit{Exfiltrate Sensitive Directory} (\(a_3\)) as goal (\(\mathcal{G}=\{a_3\}\)), thus \name{} automatically assigns it a high default goal reward of 1000 (\(r_{a_3}=1000\)).
With its default configuration (\(g=0.4\) and a base reward of 1 for all actions), the future reward calculations result in 401 for \textit{Compress Sensitive Directory} and 161 for \textit{Find Sensitive Directory}.
Driven by those rewards, \name{} executes the following sequence: \textit{Find Sensitive Directory}, \textit{Compress Sensitive Directory}, and \textit{Exfiltrate Sensitive Directory}.
Since \(a_3 \in \mathcal{G}\), the constructed sequence is a solution for the planning problem \(\mathcal{P}\).

\subsection{Coverage of Additional Tactics}
\label{subsec:emulation-of-all-tactics}

Adversary emulation should provide a wide coverage of TTPs (cf. Section~\ref{subsec:requirements}).
We designed \name{} as a plugin for Caldera to make use of Caldera's large library of techniques, covering all post-compromise tactics with over 1,700 actions in total.
In the following, we describe how \name{} enhances Caldera's capabilities with two dedicated planning components for \textit{(pre-)com\-pro\-mise} and \textit{coherent privilege escalation} techniques, respectively, to allow for the emulation of adversaries that cover pre-, initial, and post-compromise tactics.

\begin{figure}
    \centering
    \includegraphics[width=0.9\linewidth]{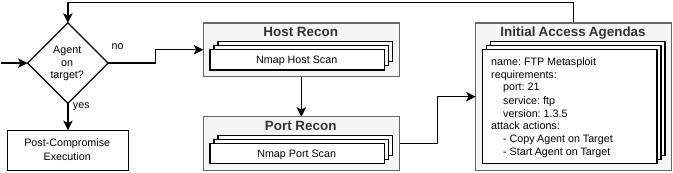}
    \caption{
      \name{} utilizes host and port scanning techniques as well as techniques for identifying and exploiting found vulnerabilities to initially compromise its target for its autonomous emulation of (pre-)compromise techniques.
    }
    \label{fig:initialcompromise}
\end{figure}

\textbf{(Pre-)Compromise.}
Many adversary emulation approaches focus on post-com\-pro\-mise tactics, often neglecting (pre-)com\-pro\-mise tactics (cf. Section~\ref{subsec:related-work}) in which adversaries intend to gain an initial foothold on the target system, e.g., by exploiting a vulnerability or delivering a payload via a phishing email, to subsequently perform their post-compromise activities.
By design, Caldera also assumes that the target system was already compromised---hence, the pre- and post-conditions of actions relate to information that was gathered on a target system.
To still facilitate the emulation of (pre-)compromise techniques, \name{} employs a separate planning component for the emulation of the three (pre-)compromise tactics, i.e., \textit{Reconnaissance}, \textit{Resource Development} and \textit{Initial Access}, which is depicted in Figure~\ref{fig:initialcompromise}.
The objective of the (pre-)compromise planning component is to start a new C2 agent on a target system, completely detached from the reward-based decision making.
At the start of an emulation, \name{} uses a dedicated C2 agent running on an adversary-controlled system.
That agent emulates reconnaissance techniques such as performing network and system scans to gather information about reachable hosts, open ports, and running services (e.g., using the network scanning tool Nmap).
Using the gathered information, \name{} identifies and exploits known weaknesses by employing pre-configured initial access agendas, i.e., short, user-defined action sequences that can compromise a target given certain requirements (cf. Figure~\ref{fig:initialcompromise}).
Finally, it compromises the target by starting a new C2 agent on it and from there continues with post-compromise actions.

\begin{figure}
    \centering
    \includegraphics[width=0.9\linewidth]{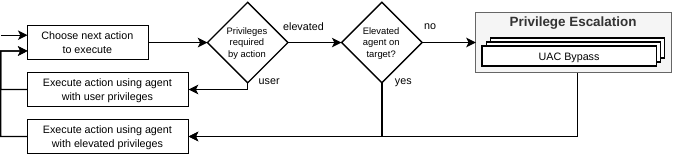}
    \caption{
      \name{} autonomously emulates coherent privilege escalation by executing privilege escalation techniques that start elevated agents on its target and executing actions in the new context of the elevated agent.
    }
    \label{fig:privilegeescalation}
\end{figure}

\textbf{Privilege Escalation.}
Real-world adversaries often switch execution contexts during their attack, e.g., using a user-level session to start an elevated session that is subsequently used to run commands.
However, many approaches (including Caldera) emulate \textit{incoherent} attacks~\cite{pruzinecKUBOFrameworkAutomated2022}, i.e., actions are executed in the emulator's context instead of the context of each other, particularly during privilege escalation.
\name{} emulates privilege escalation in a coherent manner using a dedicated planning component whose planning logic is depicted in Figure~\ref{fig:privilegeescalation}.
During an assessment, \name{} checks whether the next action to execute requires elevated privileges.
If it does and no agent with elevated privileges has yet been started on the target system, \name{} autonomously executes a privilege escalation technique to start a new elevated agent. 
Upon successfully starting the new agent, \name{} resumes the execution with the action that required elevated privileges using the new elevated agent.

\subsection{Rewards based on Action Properties}
\label{subsec:adaptable-attributes}
When running security assessments and trainings, the emulated adversary should be adaptable in its attributes, e.g., its level of expertise (cf. Section~\ref{subsec:requirements}).
We designed \name{}'s future reward calculation (cf. Section~\ref{sec:planningmethod}) to facilitate the usage of custom parameters that utilize certain properties of actions.
Defining properties of attack actions (e.g., their detectability) allows for an emulation of adversaries with desired attributes.
Furthermore, this allows implementing binary conditions based on actions' properties to facilitate the emulation of adversaries that strictly use actions with selected properties (e.g., they are used by a designated threat actor).

To give an understanding of how \name{}'s future reward calculation employs the properties of actions, we give an example of how it facilitates varying levels of difficulty for trainees by employing the \textit{detectability} of attack actions, i.e., an indicator how easy an action can be detected.
Since all publicly available threat detection products and rulesets can also be used by adversaries, they can test to which degree their actions can be detected by defenders before running an attack~\cite{uetzYouCannotEscape2023}.
Executing actions in test environments allows monitoring generated log data and alerts to determine the detectability of each action.
To represent this behavior, we extend \name{}'s future reward calculation (cf. Section~\ref{sec:planningmethod}) by a new factor \(d(a)\), i.e., the detectability of action \(a\).
Furthermore, we introduce the \textit{detectability\_weight} parameter \(w\) for configuring the influence of the detectability on the final reward (cf. Section~\ref{sec:experimentattributes}).
The \textit{adapted} future reward of an action \(f^*\) is then calculated using the following Formula 2.
For an example on how \name{} employs action properties, we refer to Experiment 2 in Section~\ref{sec:experimentattributes}.
\begin{equation}
    f^*(a) = f(a) \times\ d(a)^{w}.
\end{equation}
\subsection{Probabilistic Action Selection}
\label{subsec:attack-paths}
When repeating emulation scenarios, the emulated adversary should be able to generate varying attack paths over multiple runs, to enhance the training experience for security personnel for whom the detection of emulated attacks could be trivial when re-running scenarios (cf. Section~\ref{subsec:requirements}).
To facilitate varying attack paths that employ different actions to reach the defined goal, \name{} can be configured to employ a probabilistic decision making method.
It utilizes a weighted-random selection mechanism, wherein the weights of individual actions are set to their normalized future reward values to directly reflect the values' ratios.
As a result, \name{} generates attack paths wherein the chosen actions as well as the execution order of actions to achieve a given goal can vary.
Furthermore, \name{} can choose and execute actions that are not related to the defined goal (e.g., to obfuscate its goal) and users can configure \name{} to have multiple goals so that the outcome of the assessment is not predetermined.
To ensure reproducible randomness, \name{}'s weighted-random selection algorithm can be seeded.

%% file: sections/40-evaluation/evaluation.tex
\section{Evaluation}
\label{sec:evaluation}

We designed \name{} to enhance security assessments by employing autonomous, reward-driven decision-making with coverage of pre-, initial, and post-compromise tactics, reward calculation based on properties of actions, and optional randomness.
To evaluate our approach, we implemented it as a plugin for the popular adversary emulation platform Caldera and assessed its capabilities against the requirements established in Section~\ref{subsec:requirements} by running three exemplary experiments: Experiment 1 addresses R1 and R2, Experiment 2 addresses R3, and Experiment 3 addresses R4.
For each of the experiments, we present the test environment, \name{}'s configuration, and results.
Additionally, we discuss \name{}'s applicability and versatility in additional scenarios as well as its computational performance.

\subsection{Experiment 1: Kerberos Golden Ticket Attack on Active Directory Domain (R1, R2)}
\label{subsec:experiment-1}

\begin{figure*}
    \centering
    \includegraphics[width=1.0\linewidth]{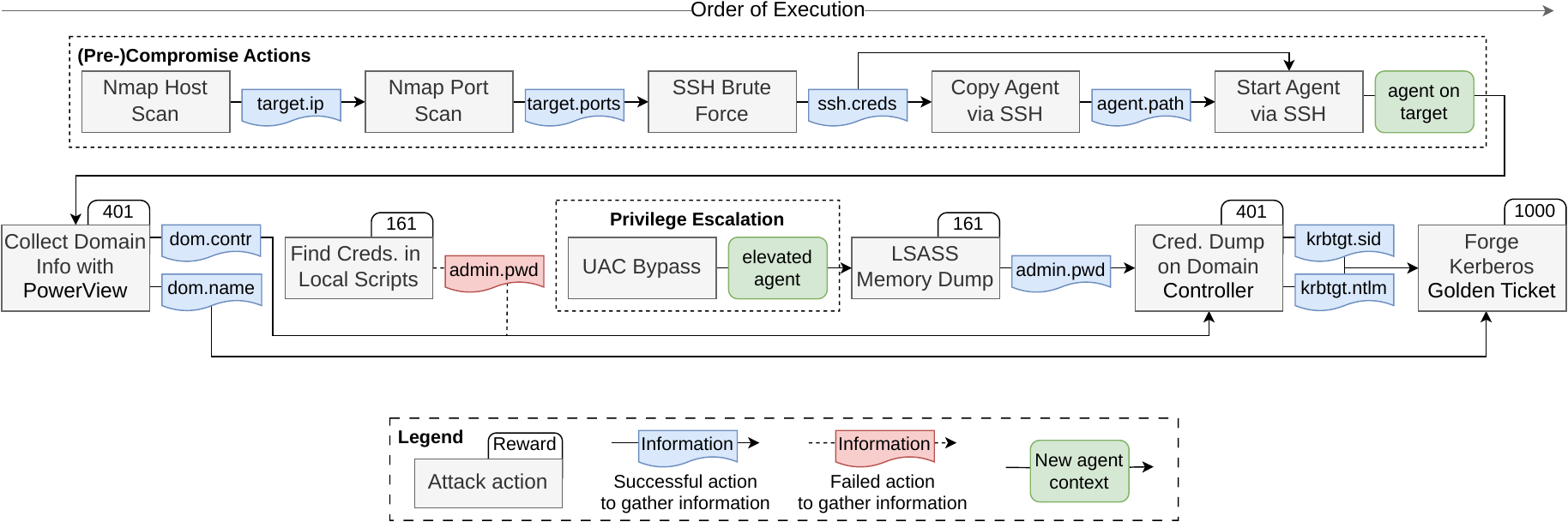}
    \caption{
      \name{} autonomously emulates the compromise of an Active Directory domain using a Kerberos Golden Ticket attack without pre-written playbooks or any prior knowledge about its target by gathering information required to execute its given goal action. The emulation includes the initial compromise of the target, coherent privilege escalation, and handling failing attack actions.
    }
    \label{fig:experiment-1-goal-pursuing}
\end{figure*}

To evaluate \name{}'s autonomous goal pursuit (R1) and TTP coverage (R2), we task it to compromise a Windows Active Directory domain using a Kerberos Golden Ticket attack~\cite{qatinahKerberosProtocolSecurity2024}.
We provide \name{} with over 1,700 attack actions, i.e., Caldera's built-in actions as well as 22 custom actions derived from an APT29 campaign in the MITRE ATT\&CK Evaluations adversary emulation library~\cite{AELAPT29}, including actions required to achieve the given goal.
We chose to implement and run the multi-step APT29 campaign instead of using Caldera's built-in techniques to assess the level of complexity \name{} can emulate.
To test active initial compromise, we replace the campaign’s original phishing-based breach with an SSH brute-force attack.
The target network includes two Windows 10 workstations and a Windows Server 2019 domain controller, configured according to the infrastructure setup instructions for the chosen campaign~\cite{AELAPT29}, including deliberately weak settings such as disabling Microsoft Defender Antivirus.
Except for an initial IP address range to scan, \name{} has no knowledge about its target before the execution of the experiment and its configuration defines only the step \textit{Forge Kerberos Golden Ticket} as goal:

\hspace*{\fill}
\begin{minipage}{0.7\textwidth}
\begin{minted}[bgcolor=LightGray, fontsize=\small, baselinestretch=0.6]{yaml}
name: Experiment 1
description: Kerberos Golden Ticket Attack on Active Directory Domain
goal_actions:
  - <forge-kerberos-golden-ticket>
\end{minted}
\end{minipage}
\hspace*{\fill}

\textit{Procedure.}
Figure~\ref{fig:experiment-1-goal-pursuing} shows the action sequence generated by \name{} using its default deterministic planning.
\name{} starts with the emulation of reconnaissance techniques, performing an Nmap host scan, followed by an Nmap port scan of the discovered hosts.
Based on the results and the presence of an open SSH port, \name{} performs an SSH brute force attack, gathering valid SSH credentials.
Those are then used to copy and start a new agent on the target, concluding the initial compromise.
\name{} then iteratively selects the highest future-reward action whose pre-conditions are satisfied to reach its goal.
It first executes \textit{Collect Domain Info with PowerView} (future reward of 401, derived from the goal action's reward), gathering the required domain name (\textit{dom.name}) and domain controller hostname (\textit{dom.contr}).
Next, it tries to satisfy the \textit{admin.pwd} precondition of \textit{Credential Dump on Domain Controller} via \textit{Find Credentials in Local Scripts}, which fails to find credentials.
\name{} then selects \textit{LSASS Memory Dump} (reward 161) as an alternative action to obtain \textit{admin.pwd}.
Since this action requires elevated privileges, \name{} initiates a privilege escalation to start a new elevated agent on the target (cf. Section~\ref{subsec:emulation-of-all-tactics}) using \textit{UAC Bypass}.
The memory dump is then executed in this new agent's context to successfully gather the required domain admin password.
Now, the pre-conditions of \textit{Credential Dump on Domain Controller} are fulfilled, allowing its execution to gather the required credential information of the Kerberos Service Account (\textit{krbtgt.sid} and \textit{krbtgt.ntlm}). With all pre-conditions of the goal action fulfilled, \name{} can finally execute \textit{Forge Kerberos Golden Ticket}, concluding the assessment.

\textit{Conclusion.}
In the conducted experiment, \name{} successfully compromises an Active Directory domain via a Kerberos Golden Ticket attack without prior target knowledge beyond an IP range.
It demonstrated that it can handle uncertainties (e.g., failed actions) by finding alternative actions to satisfy unmet pre-conditions.
By using three different agents over the course of the emulated attack, \name{} shows coherent attack sequences including correct execution context.
The attack covers pre- and initial-compromise phases and, building on Caldera’s post-compromise capabilities (cf. Section~\ref{subsec:aae}), shows that \name{} can emulate techniques of pre-, initial-, and post-compromise tactics.
The results of this experiment show that \textit{\name{} implements autonomous goal pursuit of user-defined goals without pre-written attack paths and prior knowledge of the target system while handling uncertainties (R1) and emulates techniques covering all 15 MITRE ATT\&CK tactics (R2)}.

\subsection{Experiment 2: Selective Emulation of Stealthy and Detectable Behavior (R3)}
\label{sec:experimentattributes}

\begin{figure}
    \centering
    \includegraphics[width=0.8\linewidth]{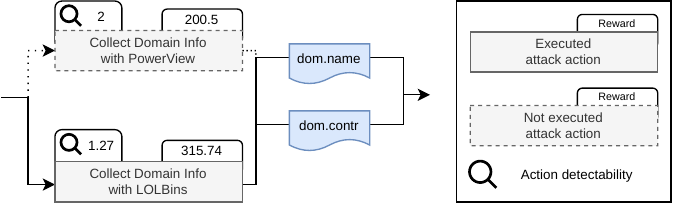}
    \caption{
        \name{} supports prioritizing actions that are harder (or deliberately easier) to detect to allow the emulation of adversaries with varying stealthiness, e.g., to conduct training exercises that match the proficiency of the trainees.
    }
    \label{fig:experiment-2-attributes}
\end{figure}

The second experiment is a variant of the first experiment and evaluates \name{}'s capabilities regarding the emulation of adversaries with adaptable attributes (R3).
We exemplarily task \name{} with the emulation of adversaries with selectively stealthy vs. detectable actions.
To provide \name{} with alternatives of actions leading towards its defined goal, we implemented a more stealthy alternative to \textit{Collect Domain Info with PowerView} that uses living-off-the-land binaries (LOLBins)~\cite{barr-smithSurvivalismSystematicAnalysis2021} instead of the well-known PowerShell script PowerView to gather the required domain information (cf. Figure~\ref{fig:experiment-2-attributes}).
We use detectability values for the two described actions---a high detectability of 2 for the original action using PowerView and a relatively low detectability of 1.27 for the alternative using LOLBins.
We provide an exemplary calculation for the actions' detectability values based on the volume of generated log data and raised alerts in a controlled test environment in Appendix~\ref{appendix:property_values}.
The configuration for this experiment includes the detectability weight that determines whether the emulated adversary should be easy or hard to detect:

\hspace*{\fill}
\begin{minipage}{0.57\textwidth}
\begin{minted}[bgcolor=LightGray, fontsize=\small, baselinestretch=0.6]{yaml}
name: Experiment 2
description: Selective emulation of stealthy adversaries
goal_actions:
  - <forge-kerberos-golden-ticket>
detectability_weight: -1
\end{minted}
\end{minipage}
\hspace*{\fill}

\textit{Procedure.}
This experiment focuses on the two described actions, therefore we omit the discussion of the other actions.
In the first experiment run, we set the detectability weight to 1, thus tasking \name{} to emulate an adversary whose actions should be easy to detect.
Using the formula from Section~\ref{subsec:adaptable-attributes}, the original action receives an adapted future reward of 802.
Due to its low detectability, the alternative action has a lower future reward of 509.27.
Thus, \name{} chooses to execute the original and easy-to-detect action using PowerView.
In the second experiment run (cf. Figure~\ref{fig:experiment-2-attributes}), we set the detectability weight to -1, thus aiming to emulate a stealthy adversary.
In contrast to the first run, now the action using LOLBins has a higher reward (315.74) than the one using PowerView (200.5).
Driven by the reward values, \name{} executes the stealthier step using LOLBins to collect the required information.

\textit{Conclusion.}
In this experiment, we selectively tasked \name{} to emulate an adversary that is either deliberately easy or hard to detect for the defending side (based on log data and alerts).
The results of the experiment show that \name{} chooses the next step to execute based on the detectability values of steps and the configured detectability weight.
Thus, we conclude that \textit{\name{} allows the targeted emulation of adversaries with adaptable attributes (R3).}

\subsection{Experiment 3: Generation of Varying Attack Paths over Multiple Runs (R4)}
\label{sec:experimentattacksequences}

To evaluate \name{}'s capabilities regarding the generation of varying attack paths when repeating a scenario to enable challenging training experiences (R4), we define another variant of the first experiment.
Our third experiment focuses on the variation of generated attack paths regarding the four main actions performed to achieve the goal.
The only change to the first experiment is enabling \name{}'s probabilistic action selection (cf. Section~\ref{subsec:attack-paths}).
We performed two instances of this experiment with the default goal reward of 1,000 (denoted as E1k) and a higher reward of 10,000 (denoted as E10k), to evaluate the influence of the goal reward value on the generated sequences.
\name{}'s configuration for E10k includes the probabilistic action selection and a custom goal reward value:

\hspace*{\fill}
\begin{minipage}{0.67\textwidth}
\begin{minted}[bgcolor=LightGray, fontsize=\small, baselinestretch=0.6]{yaml}
name: Experiment 3 - E10k
description: Generation of Varying Attack Paths over Multiple Runs
goal_actions:
  - <forge-kerberos-golden-ticket>
weighted_random: True
goal_reward: 10000
\end{minted}
\end{minipage}
\hspace*{\fill}

\begin{figure*}
\centering
\begin{subfigure}{0.49\textwidth}
    \centering
    \includegraphics[width=\linewidth]{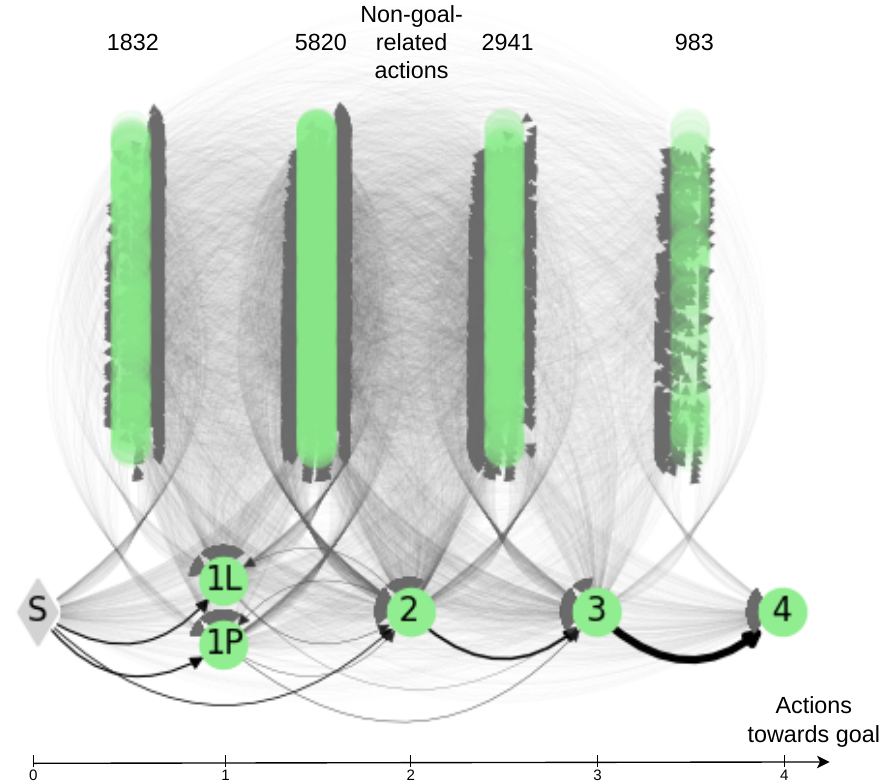}
    \caption{E1k shows a great variety in attack path generation.}
    \label{fig:sub1}
\end{subfigure}
\hfill
\begin{subfigure}{0.49\textwidth}
    \centering
    \includegraphics[width=\linewidth]{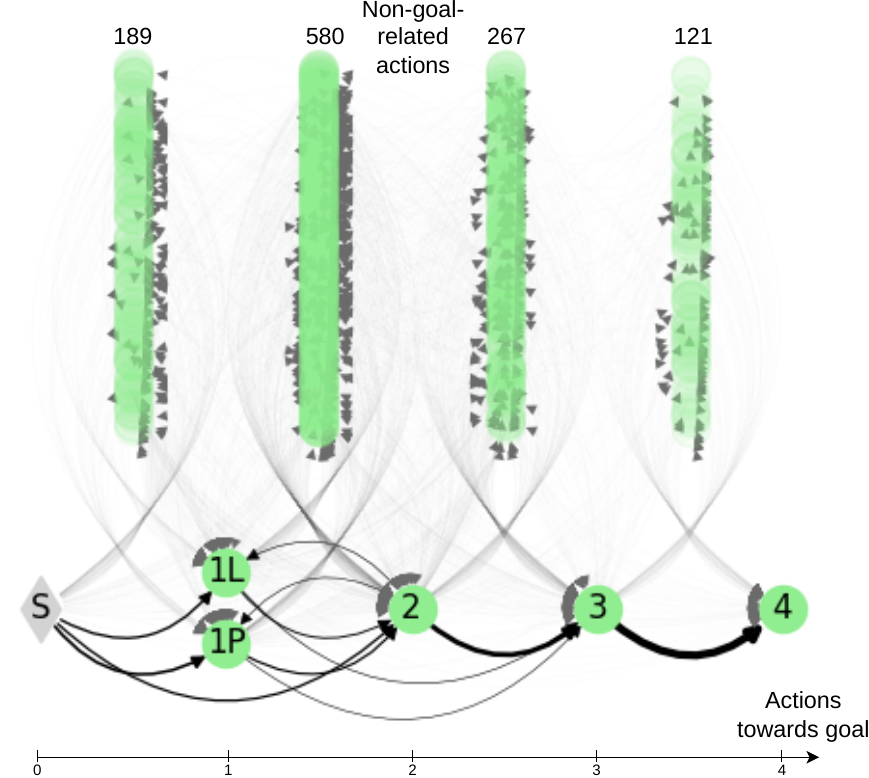}
    \caption{E10k's generated sequences are more goal-oriented.}
    \label{fig:sub2}
\end{subfigure}
\caption{
    Over the 1,000 runs for both experiment instances, respectively, \name{} generates varying attack paths using the four actions (and one action alternative) to reach the user-defined goal without any changes to its configuration.
    The strictness towards the goal depends on its configured goal reward value.
    Each non-goal-related action is shown only once, at the right-hand side of the goal-related action after which it was first executed in the 1,000 runs.
}
\label{fig:simulation}
\end{figure*}

\textit{Procedure.}
For both instances, we repeated the emulation process 1,000 times.
First, we ran E1k with the default goal reward value 1,000 times with the generated sequences shown in Figure~\ref{fig:simulation}(a).
In 1,000 runs \name{} generated 999 unique sequences (i.e., only one duplicate was generated) with an average length of 15.58 actions, the longest sequence of 57 actions, and the shortest sequence including only goal-oriented actions with a length of four.
Next, we ran E10k 1,000 times to evaluate the influence of the increased goal reward on the generated sequences (cf. Figure~\ref{fig:simulation}(b)).
In 1,000 runs \name{} generated 628 unique sequences with an average length of 5.16, the longest sequence of 12 actions, and the shortest sequence also including only goal-oriented actions with a length of four.
We can see that using a higher goal reward results in shorter, more goal-oriented sequences that include fewer non-goal-related actions.
Thus, the goal reward influences the strictness of the emulated adversary towards the defined goal.

\begin{figure}
    \centering
    \includegraphics[width=0.9\linewidth]{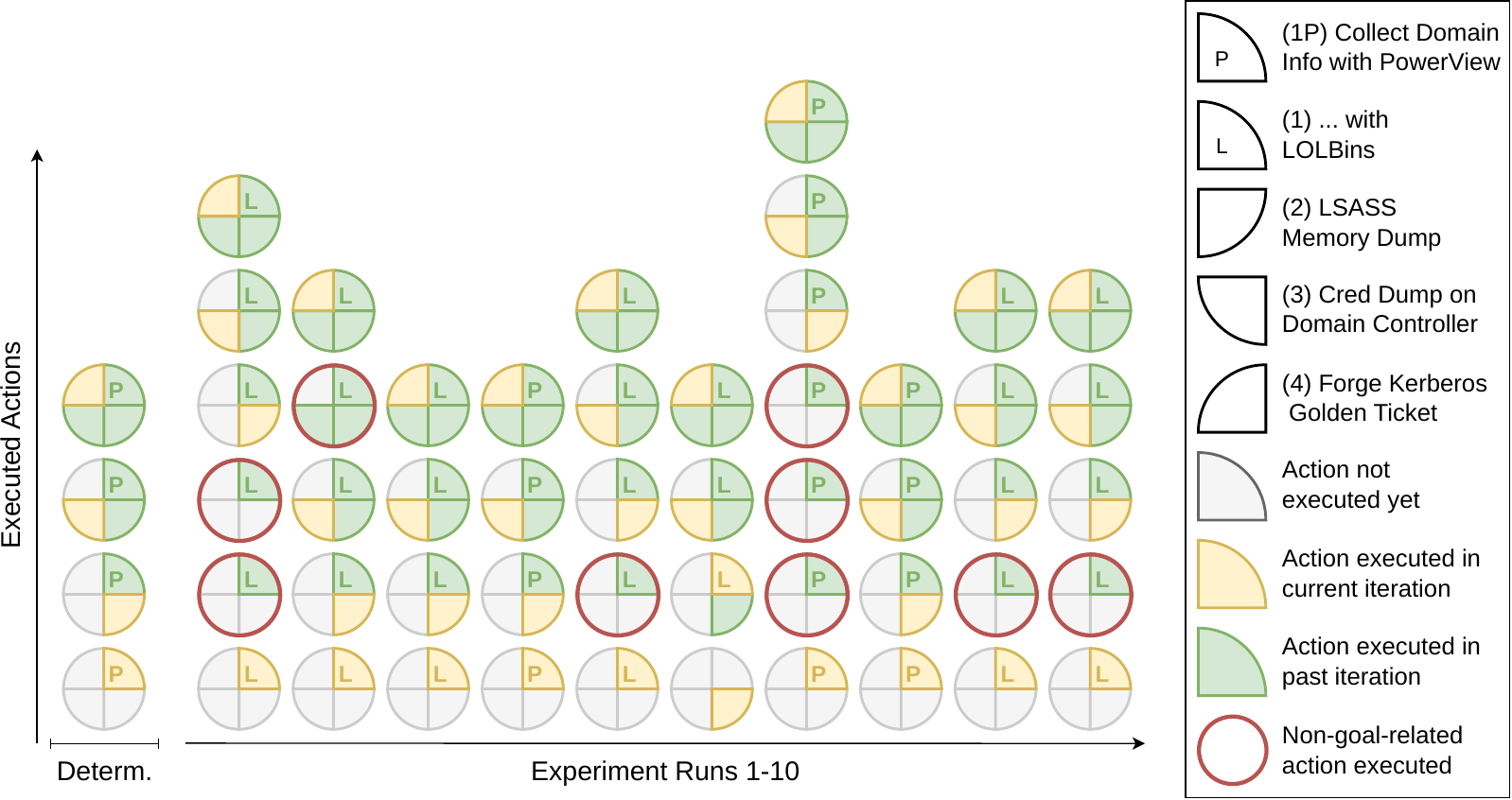}
    \caption{
      \name{} generates varying attack paths towards the objective when using its probabilistic action chaining, varying the number, order, and variants of executed actions towards the goal without any changes to its configuration.
    }
    \label{fig:experiment-3-sequences}
\end{figure}

Finally, we analyzed the attack path variety in more detail using the results of the first 10 runs of E10k as visualized in Figure~\ref{fig:experiment-3-sequences}.
As a baseline, we used the attack sequence generated by \name{} using its deterministic behavior in Experiment 1 with the generated sequence shown in the bottom row of Figure~\ref{fig:experiment-3-sequences}.
The order of executed actions is (1P), (2), (3), and finally the goal (4), without any actions that are not related to the goal.
As an example for a non-deterministically generated sequence, we describe the executed actions during the first run of E10k shown in the row above the baseline in Figure~\ref{fig:experiment-3-sequences}.
As first step, \name{} executed the LOLBins variant of the first action (1L), followed by two non-goal-related actions.
Next, actions (2) and (3) were executed as fourth and fifth steps.
After gathering all required information, \name{} finally executed the goal action (4), concluding the first run.
We note that during this experiment only actions (1) and (2) can be chained interchangeably, since action (3) requires information from actions (1) and (2) and action (4) requires information from actions (1) and (3) (cf. Figure~\ref{fig:experiment-1-goal-pursuing}).

\textit{Conclusion.}
When repeating the described experiment instances 1,000 times each without changing \name{}'s configuration, we can see that the generated sequences vary in number, selected variants, and order of executed actions.
Additionally, we see that the configurable goal reward value influences the strictness towards the defined goal and thus the length of the generated attack sequences, allowing the targeted emulation of adversaries that either work strictly towards their goal or include non-goal-related actions in their attack.
When examining runs with the same number of executed actions (e.g., Runs 4 and 6 from Figure~\ref{fig:experiment-3-sequences}), we can see that the chosen path towards the goal (i.e., the variants of actions) and the order of executed actions varies as well.
To conclude, \textit{\name{} can generate varying attack paths regarding the chosen actions, their order, and the total length of the generated path towards the same goal, e.g., for more diverse training experiences (R4).}

\subsection{Additional Scenarios}
\label{sec:generalizability}

\begin{figure*}
\centering
\begin{subfigure}{1.0\textwidth}
    \centering
    \includegraphics[width=0.9\linewidth]{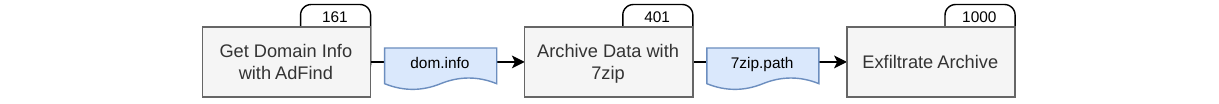}
    \caption{Attack actions from the additional FIN6 scenario with the goal to exfiltrate domain information collected with AdFind.}
    \label{fig:fin6}
\end{subfigure}
\hfill
\begin{subfigure}{1.0\textwidth}
    \centering
    \includegraphics[width=0.9\linewidth]{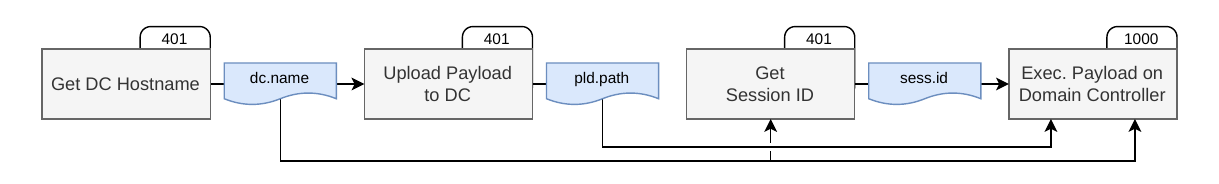}
    \caption{Attack actions from the additional APT29 scenario with the goal to move laterally to a secondary victim machine.}
    \label{fig:apt29day1}
\end{subfigure}
\caption{The additional scenarios based on the MITRE ATT\&CK Evaluations demonstrate \name{}'s broad applicability and versatility in simulating attack scenarios with diverse objectives and techniques.}
\label{fig:additional_scenarios}
\end{figure*}

We have shown that \name{} fulfills the established requirements from Section~\ref{subsec:requirements} using variations of an exemplary experiment.
To demonstrate its broad applicability and effectiveness in varying real-world scenarios allowing for the emulation of diverse adversarial behaviors, we implemented two additional scenarios from the MITRE ATT\&CK Evaluations as well as two additional initial access vectors that target Linux machines:

(1) Based on an operation of the threat group FIN6~\cite{AELFIN6}, we implemented a scenario wherein the emulated adversary is tasked to exfiltrate sensitive domain information.
This scenario includes three custom actions that employ the command-line tool AdFind for gathering the information, 7zip for compressing it, and the SSH utility tool Plink to exfiltrate the compressed archive to an attacker-controlled system (cf. Figure~\ref{fig:fin6}).
(2) We implemented an additional scenario based on another APT29 operation~\cite{AELAPT29}, wherein the attacker is tasked to move laterally through the target network to compromise a domain controller.
The scenario includes four custom actions that cover discovery of hostnames and session IDs, tool transfer from an attacker-controlled WebDAV server, and remote execution using the well-known Sysinternals Suite (cf. Figure~\ref{fig:apt29day1}).
(3) We implemented additional initial access vectors for Linux targets that exploit known vulnerabilities of a specific FTP and UnrealIRC service version, respectively, using Metasploit.

To validate that \name{} can autonomously emulate the two additional scenarios, we provided it with the described new actions as well as the over 1,700 built-in attack actions.
When configuring the respective scenario's last action as goal action, \name{} autonomously gathered the information required for the successful execution of the defined goal.
Using a vulnerable target system, we validated that \name{} can autonomously detect and exploit known vulnerabilities and weaknesses to perform the additional initial access techniques.
Thus, the additional techniques broaden \name{}'s evaluation scope, demonstrating its broad applicability and versatility in simulating attack scenarios with diverse techniques.

\subsection{Computational Performance}
After showing that \name{} fulfills the requirements from Section~\ref{subsec:requirements} and discussing its capabilities in additional scenarios, we investigate its computational performance using the setup of Experiment~1.
Using a commodity notebook (Intel Core  i7-9850H CPU @ 2.60\,GHz and 16\,GB RAM) that simultaneously hosted the experimental environment as virtual machines took \name{} approximately 12.5 minutes.
The pre-compromise and initial compromise phase (including the performed Nmap scans) took around 3.5 minutes.
The C2 communication and execution of commands constitute more than 80\,\% of the emulation time, mainly due to the beacon timer of the Caldera agents, i.e., the idle waiting between C2 communications.
Thus, the actual planning, parsing of outputs, and managing of agents performed by \name{} and Caldera took less than 20\,\% of the total emulation time. 
Choosing the next action to execute including the future reward calculation for \textit{all} (over 1,700) available actions took \name{} approximately 15 seconds on average, which allows for real-time attack emulation and should outperform manual emulation performed by human operators.

In summary, our three exemplary experiments depicting different variants of a Kerberos Golden Ticket attack against a Windows Active Directory domain have shown that \name{} can autonomously pursue a user-defined goal without prior knowledge of its target.
The emulated attack covers pre-, initial, and post-compromise tactics, facilitates adaptable adversarial attributes, and generates varying attack paths when repeating scenarios.
The additional scenarios and initial access vectors demonstrate \name{}'s broad applicability and versatility, allowing for the emulation of diverse attack techniques.
Furthermore, its computational performance allows for real-time attack emulation.

%% file: sections/50-discussion/discussion.tex
\section{Discussion and Limitations}
\label{sec:discussion}

In the following, we discuss fundamental design decisions in \name{} and potential resulting limitations.

First, while \name{} facilitates multi-step cyberattack emulation without extensive security expertise, it \textbf{requires pre- and post-conditions of actions} for constructing the links between actions used for the autonomous, goal-pursuing emulation.
Thus, particularly when adding a new action, its pre- and post-conditions have to be defined.
However, \name{} can use Caldera's built-in actions, which already provide such conditions.
Moreover, we assume that users implementing a new attack have a sufficient understanding of it to explicitly formulate its conditions.

Similarly resulting from its usage of pre- and post-conditions, \textbf{\name{} strictly follows reward values} towards a defined goal.
E.g., when given the goal to exfiltrate sensitive data, it will aim to execute this action as soon as possible, possibly compressing a staging directory containing only a single file before exfiltrating it and thus concluding the assessment, instead of staging more files for compression and exfiltration.
However, its capabilities regarding locking actions, assigning custom rewards, and updating rewards can be used to overcome this issue, e.g., by gradually increasing the reward of the compression action for each staged file.

Furthermore, strictly following reward values might result in a premature reward-driven execution of goal actions, e.g., file exfiltration, on the initial adversary-controlled system.
We address this issue by designing \name{}'s (pre-)com\-pro\-mise completely separated from the further autonomous, reward-driven execution with \textbf{predefined initial access agendas}, which implement short action sequences to ensure stringent execution of compromise techniques. %
To maintain autonomy nonetheless, \name{} checks which agendas are applicable based on gathered information and executes them in an autonomous manner.

Offering \textbf{many configuration parameters} might result in a complex configuration, our experiments show that \name{} already performs well with the chosen default parameters.
Sophisticated configurations for different scenarios can be defined and loaded using configuration files (cf. Section~\ref{sec:evaluation} and Appendix~\ref{appendix:example-scenario}).
We also note that most of the parameters are optional and only provide additional possibilities for custom adversarial behaviors (cf. Section~\ref{sec:evaluation}).

Finally, when running an emulation using \name{}, the deployed default Caldera agents and used payloads are detected and blocked by common antivirus software such as Windows Defender.
Consequently, as for other adversary emulation tools~\cite{landauerRedTeamRedemption2024}, \textbf{\name{} requires disabling antivirus software and firewalls} or adding exceptions to them, arguably reducing the realism of the assessment.
However, the survey by Landauer et al. revealed that non-disruption by defensive measurements is among the less desired aspects for users of adversary emulation tools~\cite{landauerRedTeamRedemption2024}, implying that this limitation remains acceptable in the context of practical adversary emulation.
Furthermore, Caldera supports custom payloads and agents that are not detected or blocked and \name{} facilitates attack emulation independently of the underlying payloads and implementations.

To conclude, while we acknowledge these limitations of \name{}, we are convinced that the drawbacks of the respective design decisions are outweighed by the benefits that come with them---allowing the autonomous emulation of adversaries that achieve their given goal without prior knowledge of their target while providing a wide coverage of tactics and exhibiting desired behavioral attributes and varying attack paths.

%% file: sections/60-ethical-concerns/ethical-concerns.tex
\section{Ethical Considerations}
\label{sec:concerns}

The publication of novel adversary emulation approaches raises ethical concerns due to potential for misuse.
In the following, we discuss potential risks with respect to \name{}'s key features and how we mitigate them.

As discussed in Section~\ref{sec:planningmethod}, in its core the \textbf{autonomous, reward-driven planning} of \name{} uses well-established methods.
While the additional capabilities help generate more diverse adversaries, we believe that these features do not raise additional concerns over existing approaches.
We specifically decided to not include the APT29-related actions used for our evaluation in the publicly released repository to avoid misuse of the provided techniques.
Consequently, we ensure that the benefits regarding this key feature outweigh potential remaining risks.

By additionally \textbf{covering (pre-)com\-pro\-mise and privilege escalation}, \name{} enables autonomous compromise of a target without extensive security expertise.
While this might enable malicious intentions, we only publicly release well-known and deliberately weak proof-of-concept techniques.
Furthermore, \name{} requires disabled antivirus software and firewalls, thus drastically reducing its potential for causing harm.
However, we acknowledge the theoretical risk of misuse if its capabilities get enhanced with more sophisticated techniques and custom agents---yet, such implementations require extensive offensive cybersecurity expertise.

\name{}'s other key features regarding \textbf{adaptable adversarial attributes} and \textbf{attack path variety} were implemented towards improving assessments such as training exercises in cyber ranges.
While they allow the emulation of adversaries with varying attack paths, expertise, and stealthiness, among other attributes, \name{} does not actually increase the expertise and stealthiness of the emulated adversaries.
Instead, it adapts the emulated adversarial behavior based on already implemented techniques.
Thus, we are confident that these novel features do not raise any additional security-related concerns. 

In summary, while acknowledging the ethical concerns surrounding automated adversary emulation, we believe that the benefits of publishing \name{} for the purpose of improving cybersecurity in organizations and offering a platform for academic research through more time-efficient, comprehensive, and adaptable assessments substantially outweigh the remaining risk of potential misuse.

%% file: sections/90-conclusion/conclusion.tex
\section{Conclusion}
\label{sec:conclusion}

In this work, we propose \name{}, a novel planning method for adversary emulation that aims to improve the degree of automation, comprehensiveness, and variability of security assessments.
It is designed to autonomously emulate multi-faceted adversarial behaviors to reach user-defined goals without prior knowledge of the target system while providing a wide coverage of tactics.
Its ability to autonomously navigate through various attack phases from reconnaissance over initial compromise to elevating privileges and executing follow-up objectives without predefined scripts is a substantial advancement over existing approaches.
\name{}'s reward-driven, probabilistic action selection employs a reward calculation based on action properties that allows for the emulation of adversaries with specific behavioral attributes and varying attack paths across scenario repetitions.
This configurability and variety are crucial for engaging training exercises to match the proficiency of trainees and enhance training effectiveness through unpredictability in attack paths.

We provide \name{} as a plugin for the prominent adversary emulation platform Caldera, building upon its large library of over 1,700 attack actions.
By making it open-source, we aim to spur further research and development in the field of automated adversary emulation.
We are confident that \name{} will not only be used as a tool for improving the security posture of organizations and resilience to cyberattacks, but also as a platform for academic research to test new security mechanisms against a controlled, realistic environment.

%% file: sections/95-appendix/appendix.tex
\definecolor{LightGray}{gray}{0.95}

\appendix

\section{Configuration with all possible parameters}
\label{appendix:example-scenario}

\begin{minipage}{0.79\textwidth}
  \begin{minted}[bgcolor=LightGray, fontsize=\small, baselinestretch=0.6]{yaml}
name: Configuration with all possible parameters
description: This configuration shows all possible configuration parameters.
  For a description of the parameters we refer to the technical documentation.
goal_actions:
  - <exfiltrate-staging-directory>
seed: 7567
weighted_random: True
depth: 3
discount: 0.9
default_goal_reward: 10000
default_reward: 1
default_reward_update: 0
detectability_weight: 0
default_detectability_factor: 1
action_rewards:
  <find-and-stage-sens-files>: 500
locked_actions:
  - <compress-staging-directory>
reward_updates:
  <find-and-stage-sens-files>:
    <compress-staging-directory>: 1
\end{minted}
\end{minipage}

\section{Definition of Action Properties}
\label{appendix:property_values}

We do not consider the calculation of action property values, e.g., the detectability of actions, an inherent part of \name{}.
Still, we briefly summarize our empirical approach towards user-defined action properties.
Generally, we suggest values \(<1\) for \emph{undesired} and \(>1\) for \emph{desired} properties, since they are factors in the actions' reward calculation.
For Experiment 2 (cf. Section~\ref{sec:experimentattributes}), we first calculated the \emph{detectability scores} of the respective actions when executing it in a test environment based on raised Sigma\footnote{https://github.com/SigmaHQ/sigma} alerts (weighted by alert level) and generated Sysmon log volume:
\(\mathrm{detectability\_score}(a) = \#\mathrm{high\_alerts}*3 + \#\mathrm{medium\_alerts}*2 + \#\mathrm{low\_alerts}*1 + \mathrm{log\_volume}/100\).
Next, we normalized the resulting scores to yield factors between \(0.5\) and \(2.0\) (i.e., the largest influence being halving and doubling respectively) with an exponential increase (determined by the denominator 10):
\(\mathrm{detectability}(a) = 2 - (1.5 / e^{\mathrm{detectability\_score}(a)/10})\).
For example, executing \emph{Collect Domain Info with LOLBins} in our test environment resulted in one high level and two medium level Sigma alerts and a total of 25 Sysmon logs.
The resulting \emph{detectability\_score} is \(1*3 + 2*2 + 0*1 + 25/100=7.25\).
Normalizing this value results in a \emph{detectability} of \(2 - (1.5 / e^{7.25/10}) = 1.27\).